\documentclass[journal]{IEEEtran}

\usepackage[utf8]{inputenc}
\usepackage[most]{tcolorbox}

\usepackage{booktabs}

\usepackage{graphicx}
\usepackage{multirow}
\usepackage{setspace}
\usepackage{amsmath}
\usepackage{amssymb} %
\usepackage{amsfonts} %
\usepackage{calligra} %
\usepackage{mathtools}
\usepackage{amsthm} %
\usepackage{hyperref}
\usepackage{array}
\usepackage{enumitem}

\usepackage{pgfplots}
\usepackage{pgfplotstable}
\usepgfplotslibrary{statistics}

\usepackage{comment}
\usepackage{listings}

\usepackage{circledsteps}

\usetikzlibrary{patterns}

\usepackage{blindtext}

\usepackage{acronym}

\usepackage{xcolor, colortbl}

\usepackage{float}
\usepackage{textcomp}

\newcommand{\change}[1]{#1}

\usepackage{algorithm}
\usepackage{algorithmic}

\usepackage{url}

\usepackage[all=normal,paragraphs=tight, floats=tight,indent=tight]{savetrees}
\usepackage{pifont}
\usepackage{framed}
\usepackage{soul}
\usepackage{csquotes}
\usepackage{multirow}
\usepackage{array}
\newcolumntype{L}[1]{>{\raggedright\let\newline\\\arraybackslash\hspace{0pt}}m{#1}}
\newcolumntype{C}[1]{>{\centering\let\newline\\\arraybackslash\hspace{0pt}}m{#1}}
\newcolumntype{R}[1]{>{\raggedleft\let\newline\\\arraybackslash\hspace{0pt}}m{#1}}
\newcolumntype{H}{>{\collectcell\lstinline}l<{\endcollectcell}}
\usepackage{url}
\MakeOuterQuote{"}
\usepackage[caption=false,font=footnotesize,labelformat=simple]{subfig} %

\input{section/preamble/preamblelistingsconf}
\acrodef{CPS}{Cyber-Physical System}
\acrodef{IoT}{Internet of Things}
\acrodef{HDL}{Hardware Description Language}
\acrodef{CAD}{Computer-Aided Design}
\acrodef{EDA}{Electronic Design Automation}
\acrodef{HPC}{High-Performance Computing}
\acrodef{DL}{deep learning}
\acrodef{ML}{machine learning}
\acrodef{NLP}{natural language processing}
\acrodef{IC}{Integrated Circuit}
\acrodef{CWE}[CWE]{Common Weakness Enumeration}
\newcommand{\ignore}[1]{{}}

\newcommand{\squishlist}{
	\begin{list}{$\bullet$}
		{ \setlength{\itemsep}{0pt}
			\setlength{\parsep}{1pt}
			\setlength{\topsep}{1pt}
			\setlength{\partopsep}{0pt}
			\setlength{\leftmargin}{0.9em}
			\setlength{\labelwidth}{1.5em}
			\setlength{\labelsep}{0.4em} } }
	\newcommand{\squishend}{
	\end{list}  }

\definecolor{graphFirst}{RGB}{2,136,209} %
\definecolor{graphSecond}{RGB}{211,47,47} %
\definecolor{graphThird}{RGB}{245,124,0} %
\definecolor{graphFourth}{RGB}{56,142,60} %
\definecolor{graphFifth}{RGB}{81,45,168} %
\definecolor{graphSixth}{RGB}{69,90,100} %
\definecolor{graphSeventh}{RGB}{251,192,45} %
\definecolor{backgroundSecond}{RGB}{239,154,154} %
\definecolor{backgroundThird}{RGB}{255,204,128} %
\definecolor{backgroundFourth}{RGB}{165,214,167} %
\definecolor{backgroundFifth}{RGB}{179,157,219} %
\definecolor{backgroundSixth}{RGB}{176,190,197} %
\definecolor{backgroundSeventh}{RGB}{255,245,157} %

\begin{document}

\title{Needle in a Haystack: Detecting Subtle Malicious Edits to Additive Manufacturing G-code Files}

\author{Caleb~Beckwith, Harsh~Sankar~Naicker \textit{(Member, IEEE)}, Svara~Mehta, Viba~R.~Udupa \textit{(Member, IEEE)}, Nghia~Tri~Nim, Varun~Gadre, Hammond~Pearce \textit{(Member, IEEE)}, Gary Mac, %
Nikhil Gupta \textit{(Senior Member, IEEE)}
\thanks{C. Beckwith is with the Dept. of Mechanical Engineering, NYC College of Technology, Brooklyn, NY 11201 USA.}%
\thanks{H. Naicker is with the School of Electronics Engineering, Vellore Institute of Technology, Kelambakkam, Chennai, Tamil Nadu 600127, India.}%
\thanks{S. Mehta is with the Dept. of Mechanical Engineering, Indian Institute of Technology Goa, Ponda, Goa 403401, India.}%
\thanks{V. R. Udupa is with the Dept. of Mechanical Engineering, National Institute of Technology Surathkal, Mangalore, Karnataka 575025, India.}%
\thanks{N. Tri Nim is with the Science Division, New York University Abu Dhabi, Abu Dhabi, UAE.}%
\thanks{V. Gadre is with the Dept. of Mechanical Engineering, Indian Institute of Technology Kanpur, Kanpur, Uttar Pradesh 208016, India.}%
\thanks{H. Pearce (corresponding author, e-mail hammond.pearce@nyu.edu) is with the Dept.
	of Electrical and Computer Engineering, New York University, Brooklyn NY 11201 USA.}%
\thanks{G. Mac and N. Gupta are with the Dept. of Mechanical and Aerospace Engineering, New York University, Brooklyn NY 11201 USA.}%
}%
\maketitle

\begin{abstract}

Increasing usage of Digital Manufacturing (DM) in safety-critical domains is increasing attention on the cybersecurity of the manufacturing process, as malicious third parties might aim to introduce defects in digital designs.
In general, the DM process involves creating a digital object (as CAD files) before using a slicer program to convert the models into printing instructions (e.g. g-code) suitable for the target printer.
As the g-code is an intermediate machine format, malicious edits may be difficult to detect, especially when the golden (original) models are not available to the manufacturer.
In this work we aim to quantify this hypothesis through a red-team/blue-team case study, whereby the red-team aims to introduce subtle defects that would impact the properties (strengths) of the 3D printed parts, and the blue-team aims to detect these modifications in the absence of the golden models.
The case study had two sets of models, the first with 180 designs (with 2 compromised using 2 methods) and the second with 4320 designs (with 60 compromised using 6 methods).
Using statistical modelling and machine learning (ML), the blue-team was able to detect all the compromises in the first set of data, and 50 of the compromises in the second.

\end{abstract}

\section{Introduction}
\label{sec:introduction}

Digital Manufacturing (DM) is increasingly used in safety-critical applications \change{across} domains such as the aerospace, automotive, medical, and military sectors~\cite{nickels_am_2015,tiwari_selection_2015}. 
Consequently, the potential \change{target space} for malicious third parties \change{is} increasing, with motivations for both espionage (information theft) and sabotage (compromising machines or model data). %

\begin{figure}[t]
    \centering
    \includegraphics[width=0.8\columnwidth]{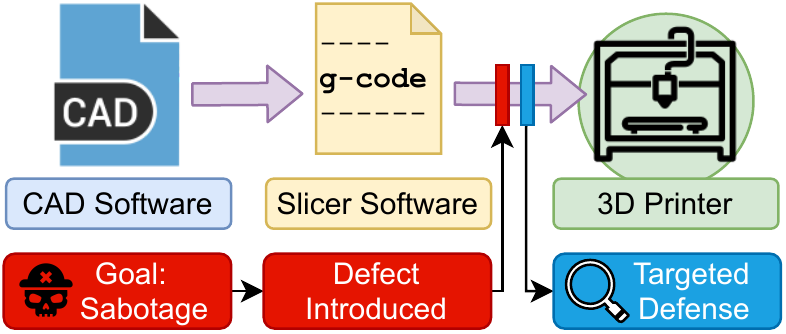}
    \caption{The red-team sabotage the \texttt{g-code} instructions prior to printing. The blue-team aims to detect the compromised files.}
    \label{fig:gcode-attack-vector}
    \vspace{-5mm}
\end{figure}

One attractive target for malicious actors in this space \change{is the \texttt{g-code} files which} detail the machine instructions used to manufacture each part~\cite{moore_vulnerability_2016,gupta_additive_2020}.
These files may contain subtractive drill/mill commands in the case of CNC-style machines, or filament extrusion commands in the case of additive manufacturing (AM).
\change{The \texttt{g-code} files are often generated externally to the machines that execute them, and due to the lossy conversions in their generation, reverse-engineering the original model files for comparison is a difficult process~\cite{tsoutsos_secure_2017}.}
\change{As such, malicious edits to \texttt{g-code} may go unnoticed.}

In this work, we thus examine the \change{design space for attacking and defending} \texttt{g-code} through a red-team/blue-team exercise (as depicted in Fig.~\ref{fig:gcode-attack-vector}).
\change{Here,} the red-team introduces defects into a sample set of \texttt{g-code} files and the blue-team detects the tampered files. 
The red-team focused on \change{mechanisms to mutate} \texttt{g-code} instructions to reduce part strength \change{in small numbers of randomly chosen files (simulating a subtle attack)}.
The blue-team then \change{proposed and utilized} 
statistical analysis and machine learning (ML) techniques as possible detection strategies \change{before a blind evaluation.} %
Overall, the red-team devised 6 different compromise methodologies, and the blue-team was able to detect 5 of these across two different datasets.

The rest of this paper is organized as follows.
Section~\ref{sec:background} covers the background and related work.
Section~\ref{sec:methodology} then describes the methodologies used by the red and blue teams.
Section~\ref{sec:evaluation} evaluates these methodologies and discusses the limitations of the approaches.
Section~\ref{sec:conclusions} then concludes.

\section{Background and Related Work}
\label{sec:background}

Before 3D CAD model files are converted to \texttt{g-code} through the use of a slicer program, they
must usually be converted to stereolithography \texttt{STL} files---imperfect representations of the original data, where curved surfaces are discretized into tessellated triangles. When converted to \texttt{g-code}, \change{these surface triangles are then encoded as lines of filament}. \change{Both conversions lose original design information and present} opportunities for malicious actors to introduce defects into parts that may go unnoticed~\cite{chen_security_2017}. 

\change{As such, just as CAD and STL files may be potential attack targets}~\cite{belikovetsky_dr0wned_2017}, \texttt{g-code} \change{files must also be considered} vulnerable. 
Example attacks have already included compromising the USB communication between PC and printer~\cite{moore_vulnerability_2016} and altering the 3D printer firmware responsible for interpreting the incoming \texttt{g-code}~\cite{pearce_flaw3d_2021,moore_implications_2017}.

While there are proposals for analyzing the \texttt{g-code} to detect introduced defects through finite element analysis simulations~\cite{tsoutsos_secure_2017}, the process for this analysis is computationally intensive and difficult, requiring derivation of the original CAD models.
As a result, other approaches have been investigated, including those from the information security domain (e.g. cryptographically signing production files to detect tampering~\cite{safford_hardware_2019}). However, depending on the attack model, tampering of the \texttt{g-code} data could occur prior to the signing of the files \change{or after a verification has occurred. }

Likewise, detecting compromised \texttt{g-code} has also been suggested via \textit{side-channel analysis}: while the initial focus has been on demonstrating  information leakage~\cite{gatlin_detecting_2019,graves_characteristic_2019}, these same channels may be used for monitoring of part defect and malicious edits~\cite{belikovetsky_digital_2019,liu_online_2021}. 
Similarly, using a vision-based ML approach to detect defects during print has also been proposed~\cite{farhan_khan_real-time_2021}.
However, these run-time approaches may only be performed during the printing process, at which point valuable resources (filament, machine time) have already been expended.
As such, in this work we propose to detect defects in the \texttt{g-code} prior to the printing process.

\section{Methodology}
\label{sec:methodology}

In this \change{work} we consider compromised \texttt{g-code} in the absence of the original CAD model files for validation. This threat vector is possible in manufacturing-as-a-service (MaaS), where only the production files may be provided to the manufacturer. 
\change{The compromise could happen at any stage before manufacturing, e.g. by a malicious slicer program or from modifications by a third party transiting the data.}
A red-team/blue-team exercise was thus conducted with the two teams isolated from one another.
The red-team, consisting of the latter three authors of the paper, constructed two datasets D1 and D2 which contained many `good' (non-compromised) \texttt{g-code} files and a small number of `bad' (compromised) \texttt{g-code} files.
The blue-team, consisting of the first six authors of the paper, were tasked with isolating the `bad' models.

\subsection{The datasets}

\begin{figure}[b]
    \centering
    \includegraphics[width=0.5\textwidth]{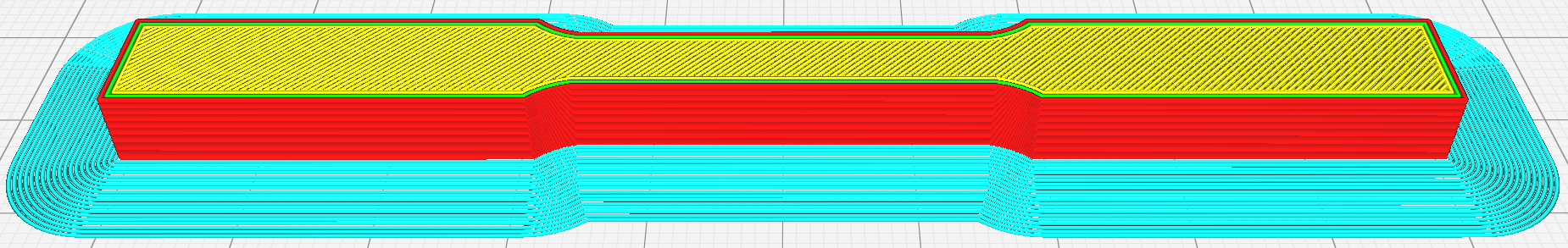}
    \vspace{-7mm}
    \caption{The control specimen for Dataset D1 viewed in Ultimaker Cura.}
    \label{fig:D1_0}
\end{figure}

\begin{figure}[t]
    \centering
    \includegraphics[width=0.3\textwidth]{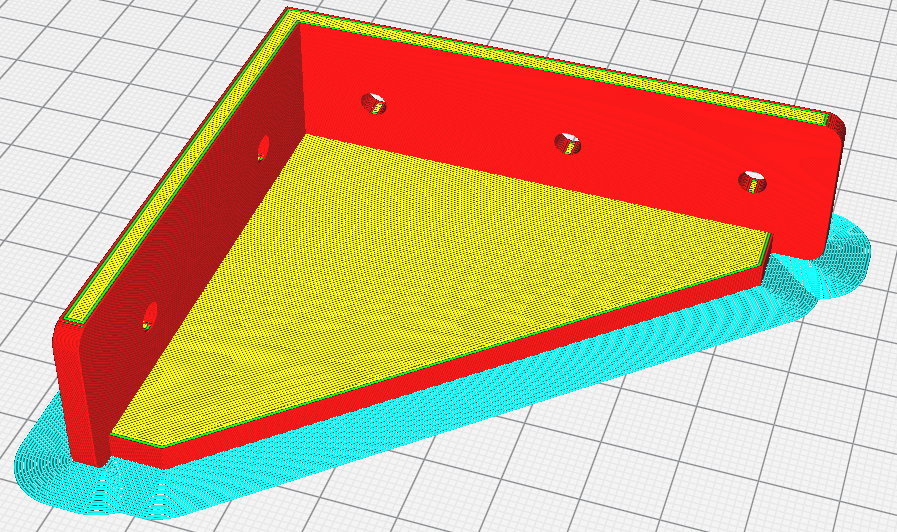}
    \vspace{-2mm}
    \caption{The control specimen for Dataset D2 viewed in Ultimaker Cura.}
    \vspace{-4mm}
    \label{fig:D2_0}
\end{figure}

To generate each dataset, we take a single STL CAD model file and perform a rotate-then-slice using Ultimaker Cura. By using a number of distinct rotations with respect to the build plate, each generated \texttt{g-code} file is unique (as the \texttt{g-code} will be optimized for that particular rotation).  
Dataset D1 consists of a tensile test specimen (Fig.~\ref{fig:D1_0}) like that in \cite{pearce_flaw3d_2021}. The CAD model is rotate-then-sliced by 1$^\circ$ 180 times in the Y direction, with slicer settings (skirt, 0.4mm nozzle, grid infill pattern, infill line distance 2mm).
Dataset D2 consists of a bracket \change{(Fig.~\ref{fig:D2_0}) which is rotated} 4320 times in total (rotate-then-slice by 0.25$^\circ$ 1440 times for each face against the print bed), with slicer settings (skirt, 0.4mm nozzle, cubic infill pattern, infill line distance 4mm).

\subsection{Red-team: Vulnerability insertion}

\begin{table}[t]
\caption{Red-team Dataset vulnerability designs. `n.D1'/`n.D2' indicates how many \texttt{g-code} files were compromised using this method in the 180-model D1 dataset / 4320-model D2 dataset respectively. `Range' indicates how much of the \texttt{g-code} file was altered - either the middle 50\,\%, or the entire 100\,\%.}
\label{tbl:defect-strategies}
\centering
\resizebox{0.9\columnwidth}{!}{%
\begin{tabular}{|l|l|l|l|L{6cm}|}
\hline
\textbf{ID} & \textbf{n.D1?} & \textbf{n.D2?} & \textbf{Range} & \textbf{Description}                                                         \\ \hline
1             & 2            & 10            & 50\,\%           & Origin~in~\cite{pearce_flaw3d_2021}. Coverts every 4th \texttt{G1} command to \texttt{G0}.                        \\  \hline
2             &              & 10            & 50\,\%           & Converts every 4th \texttt{G1} command to \texttt{G0}, and adds a \texttt{G1} `blob' extrusion.                     \\ \hline
3             &              & 10            & 100\,\%          & Origin~in~\cite{pearce_flaw3d_2021}. Reduces extrusion globally by 50\%.                         \\ \hline
4             &              & 10            & 50\,\%           & Every 4th \texttt{G1} command has extrusion value set to previous extrusion.           \\ \hline
5             &              & 10            & 50\,\%           & Every 4th \texttt{G1} command has extrusion value set to previous extrusion + 0.0001. \\ \hline
6             &              & 10            & 50\,\%           & Deletes~every~4th~\texttt{G1} line with no replacement.                               \\ \hline
\end{tabular}
}
\vspace{-5mm}
\end{table}

\begin{figure}[b]
    \centering
    \vspace{-5mm}
    \includegraphics[width=0.48\textwidth]{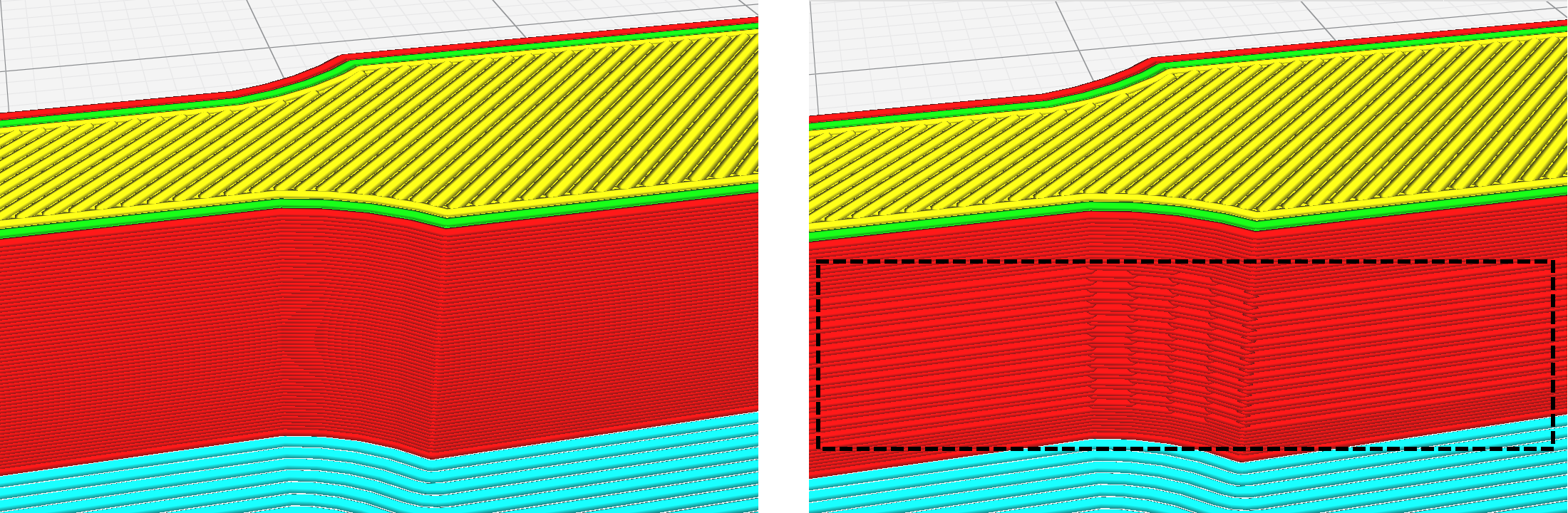}
    \vspace{-1mm}
    \caption{Comparison of good (left) verses compromised (right) \texttt{g-code} in Dataset 1 by method ID1 when viewed in Ultimaker Cura. The dotted black box indicates the area with altered print commands.}
    \label{fig:D1_compromise}
\end{figure}

In this \change{work, the compromises were designed similarly to the methodology for reducing the strength of printed models presented} in \cite{pearce_flaw3d_2021}. 
Six different vulnerability strategies were devised, and are listed in Table~\ref{tbl:defect-strategies}.
They function primarily by altering the \texttt{g-code} \texttt{G1} (linear move and extrude) command to introduce gaps (voids) in the printing process by subverting the absolute frame of reference used within common \texttt{g-code}.
As an example, 
consider three back-to-back commands \texttt{(G1 X1 Y1 E1)}, \texttt{(G1 X5 Y1 E2)}, and \texttt{(G1 X5 Y5 E3)}. The total extruded material is 1mm after the first command, 2mm after the second (it extrudes 1mm), and 3mm after the third (it also extrudes 1mm). 
If the Trojan alters the second command to \texttt{(G0 X5 Y1)}, no material is extruded along the same route. 
Instead, the third command will now extrude 2mm, such that the total material used remains 3mm.

As such, %
ID1 and ID2 convert some \texttt{G1} commands to \texttt{G0} (linear move) commands, dropping the extrusions. ID2 adds an additional `blob' extrusion (keeping the head in place and depositing the missing material) at the end of the original move.
ID3 changes all \texttt{G1} commands to use 50\,\% less filament.
ID4 and ID5 set some of the \texttt{G1} extrusions to negligible values, essentially making them \texttt{G0} commands. 
Finally, ID6 simply deletes some of the \texttt{G1} commands. 
Fig.~\ref{fig:D1_compromise} depicts ID1 on D1---with the middle 50\,\% of the part defective (lacking extrusions).

\subsection{Blue-team: Vulnerability detection}

\begin{figure}[t]
    \centering
    \includegraphics[width=0.4\textwidth]{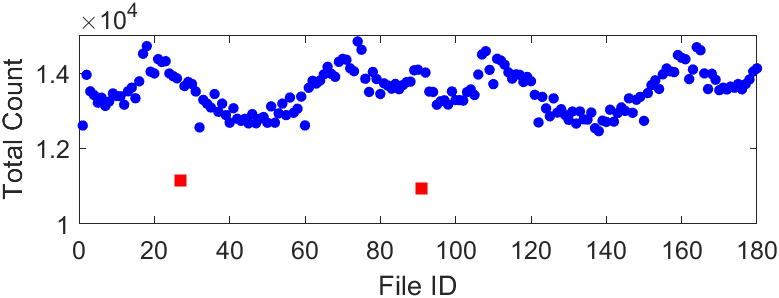}
    \caption{Distribution of \texttt{G1} commands in Dataset D1. Outliers are in red.}
    \label{fig:D1_statistics_G1}
    \vspace{-3mm}
\end{figure}
\begin{figure}[t]
    \centering
    \includegraphics[width=0.4\textwidth]{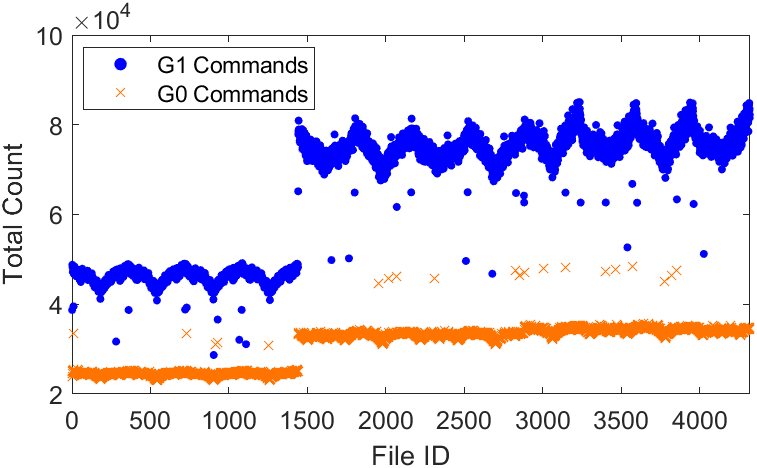}
    \caption{Distribution of \texttt{G1} and \texttt{G0} commands in Dataset D2.}
    \label{fig:D2_statistics_G0_G1}
    \vspace{-3mm}
\end{figure}

\subsubsection{Statistical Analysis}
The first strategy utilized by the blue-team was to perform a statistical feature extraction of the \texttt{g-code} files. These included the number of layers, the boundaries of the values of \texttt{X}, \texttt{Y}, and \texttt{Z} commands, and the material length extruded. Given that the two most common \texttt{g-code} commands are \texttt{G0} and \texttt{G1}, the blue-team reasoned that any compromise would likely effect the statistical properties of these commands, and so they also included the number of the \texttt{G0} / \texttt{G1} commands. %
These features were then examined using \texttt{matplotlib} and \texttt{seaborn} in Python to visually identify outliers. As can be seen in Fig.~\ref{fig:D1_statistics_G1}, this strategy appeared to identify the two files with anomalous counts. Manual inspection of the \texttt{g-code} in Ultimaker Cura (e.g. Fig.~\ref{fig:D1_compromise}) then confirmed these files were defective.
This `Basic Statistical Analysis' strategy was thus attempted over the larger Dataset D2. However, again using Cura, the blue-team determined that this single-feature approach was now returning false positives.

As such, the outlier detection was expanded to include both \texttt{G1} and \texttt{G0}, as depicted in Fig.~\ref{fig:D2_statistics_G0_G1}.
From here it can be concluded that files may have either `too few' \texttt{G0} commands, `too many' \texttt{G1} commands, or both. From \change{this,} it was possible to identify a total of 30 files \change{that} seemed to have defects when viewed in Cura.
A further anomaly was identified when considering the extrusion values \texttt{E}. \change{The} vast majority of \change{the} files \change{had }5 decimal points, however, some files had \texttt{E} commands with more or fewer decimals. These files were thus determined to be corrupted creating two distinct categories: Files that were given extra extrusion distance; and files that were given less extrusion distance. For each category, 10 more files were found for a total of 50 files. The blue-team termed this complete analysis the `Combined Statistical Analysis'.
 
\subsubsection{Machine learning (ML) based approach}

\begin{figure}[t]
\centering
\includegraphics[width=\columnwidth]{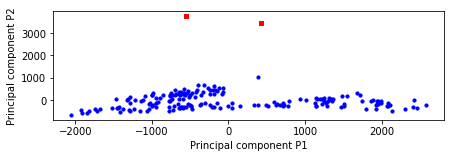}
\vspace{-5mm}
\caption{Dataset D1 outlier detection using Principle Component Analysis and Agglomerative Clustering. Outliers in red.}
\label{fig:D1_PC}
\end{figure}

\begin{figure}[t]
\centering
\includegraphics[width=\columnwidth]{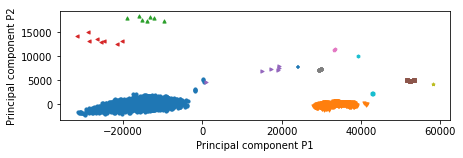}
\vspace{-5mm}
\caption{Dataset D2 outlier detection using Principle Component Analysis and Mean Shift Algorithm. \change{The two major clusters are the blue circles and orange down-arrows. Other cluster colors/shapes} indicate outlier groups.}
\label{fig:D2_PC}
\end{figure}

\begin{figure}[t]
    \centering
    \includegraphics[width=0.47\textwidth]{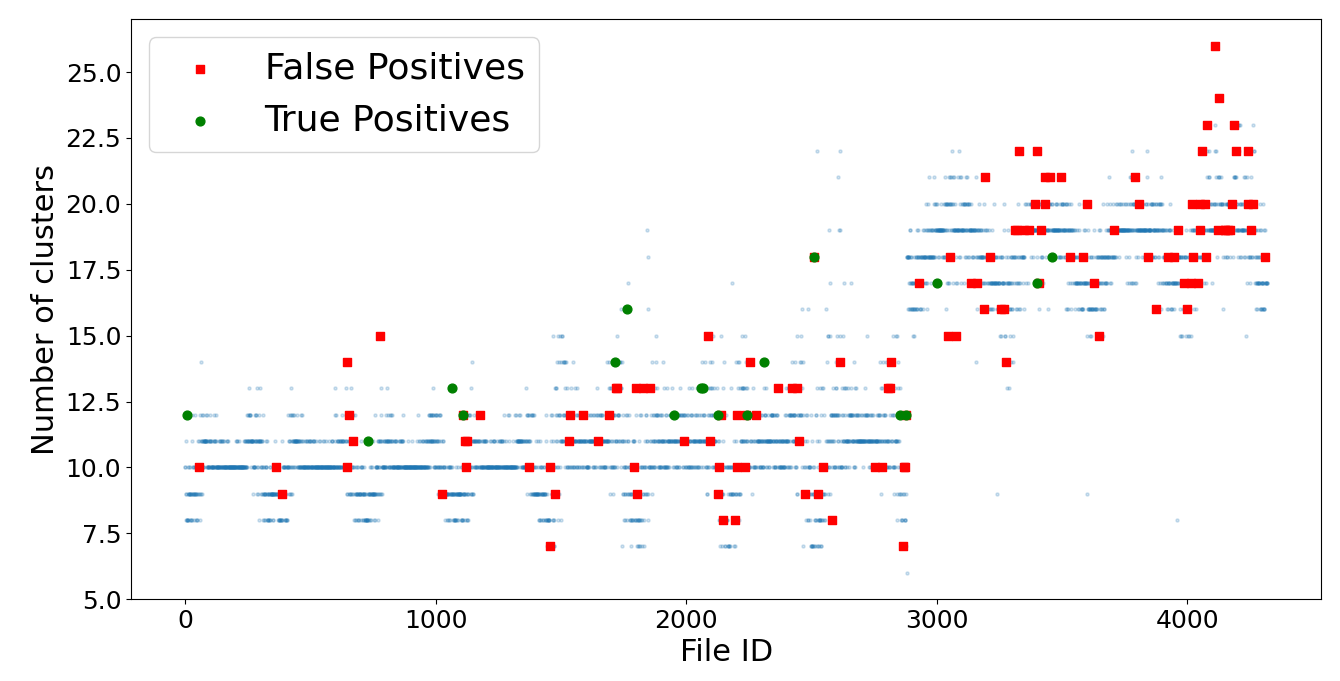}
    \vspace{-5mm}
    \caption{Plot of the number of clusters using DBSCAN. Detected outliers are marked with \change{green circles/red squares} depending on their final correctness (not provided to blue-team).}
    \label{fig:clusters}
    \vspace{-5mm}
\end{figure}

The first ML-based method examined for identifying outliers involved using Principal Component Analysis (PCA) followed by various clustering algorithms. %
Here, the \texttt{g-code} files were converted to data frames using \change{the}
 \texttt{gcodeparser} package~\cite{everitt_gcodeparser_2021}. These data frames were then analyzed to produce a count of each command type along with the total count of lines present in the file. %
PCA was then utilized to reduce the dimensions of each data frame from 11 to 2.
Following this, various clustering algorithms like agglomerative clustering, mean shift algorithm, etc. were used and scatter plots were obtained of the clusters for each dataset D1 (Fig.~\ref{fig:D1_PC}) and D2 (Fig.~\ref{fig:D2_PC}). Here, although the outliers from D1 were again identified, the D2 outliers were less consistent with the previously identified defective files.

An alternative approach was thus devised using clustering-based unsupervised learning, specifically, ``density-based spatial clustering of applications with noise'' (DBSCAN). Here, the earlier data frames were clustered in Python using \texttt{scikit}'s DBSCAN to produce Fig.~\ref{fig:clusters}. Here, outliers are represented as spikes on the graph. However, while this method had some overlap with the results from the statistical approach, there were also other files identified.

\section{Evaluation}
\label{sec:evaluation}

\begin{table}[]
\caption{Overall detection results}
\label{ref:tbl-results}
\centering
\resizebox{\columnwidth}{!}{%
\begin{tabular}{l|l|l|l|l|l|l|l|l|}
\cline{2-9}
                                                   & \multicolumn{4}{c|}{\textbf{Dataset D1}}                                                                                                       & \multicolumn{4}{c|}{\textbf{Dataset D2}}                                                                                                          \\ \hline
\multicolumn{1}{|l|}{\textbf{Method}}              & \multicolumn{1}{c|}{\textbf{T.P.}} & \multicolumn{1}{c|}{\textbf{F.P}} & \multicolumn{1}{c|}{\textbf{T.N}} & \multicolumn{1}{c|}{\textbf{F.N}} & \multicolumn{1}{c|}{\textbf{T.P.}} & \multicolumn{1}{c|}{\textbf{F.P.}} & \multicolumn{1}{c|}{\textbf{T.N.}} & \multicolumn{1}{c|}{\textbf{F.N.}} \\ \hline
\multicolumn{1}{|l|}{\textit{(Correct Value)}}                & \textit{2}                         & \textit{0}                        & \textit{178}                      & \textit{0}                        & \textit{60}                        & \textit{0}                         & \textit{4260}                      & \textit{0}                         \\ \hline \hline
\multicolumn{1}{|l|}{Single Statistical Analysis}         & 2                                  & 0                                 & 178                               & 0                                 & 29                                 & 21                                  & 4260                               & 10                                 \\ \hline
\multicolumn{1}{|l|}{Combined Statistical Analysis}         & 2                                  & 0                                 & 178                               & 0                                 & 50                                 & 0                                  & 4260                               & 10                                 \\ \hline
\multicolumn{1}{|l|}{PCA and Clustering} & 2                                  & 0                                 & 178                               & 0                                 & 28                                  & 7                                  & 4232                                  & 32                                  \\ \hline
\multicolumn{1}{|l|}{DBSCAN}                       & 1                                  & 0                                 & 178                                 & 1                                 & 35                                 & 7                                  & 4253                               & 25                                 \\ \hline
\end{tabular}
}
\end{table}

The overall detection strategy results are presented in Table~\ref{ref:tbl-results}, with each method paired with the detection results True Positive (T.P.) (i.e. defect present in \texttt{g-code} file and algorithm correctly detected this as an outlier), False Positive (F.P.) (i.e. no defect but wrongly identified as \change{an} outlier), True Negative (T.N.) (i.e. no defect and correctly identified as not an outlier), False Negative (F.N.) (i.e. defect present but wrongly identified as not an outlier) for each of the datasets D1 and D2.
As can be seen, the different strategies proved largely successful at identifying the defects (outliers) in D1. However, none of the methods identified all defects in D2. The best approach, `Combined Statistical Analysis', was able to detect all compromised files by all defect strategies other than those compromised by ID3 (which reduced extrusion globally by 50\,\%)---indeed, none of the detection strategies detected any files by ID3. This was a surprising result, as previous work~\cite{pearce_flaw3d_2021} determined that this methodology of \texttt{g-code} compromise would be the most obvious when considering post-manufacturing checks. This is likely because the defect was global, and so none of the algorithms saw sudden changes in the files caused by the defect.
\change{Finally, while the} \change{machine-learning-based} approaches show some promise, it appears they need further refinement before they will function as well as the `Combined Statistical Analysis'. %

\change{\textbf{Limitations:} In this work the datasets were synthetically generated from the original models presented in Fig.~\ref{fig:D1_0} and~\ref{fig:D2_0}. While the \texttt{g-code} for each individual rotation was quite distinct (e.g. Fig.~\ref{fig:D1_statistics_G1} and~\ref{fig:D2_statistics_G0_G1}), it is likely that the conceptual similarities between each model aided in the detection strategies. Further, our attack model assumed that a small minority of files would be compromised by the attackers. Future work could aim to evaluate both of these cases---(1), where datasets include models with unrelated geometries, and (2) some datasets where a majority or all files are compromised.}

\section{Conclusions}
\label{sec:conclusions}
Automatically detecting defects in \texttt{g-code} is an important step towards the cybersecurity of Manufacturing-as-a-Service (MaaS) production processes.
Here, models may be required to be validated without access to the original \change{design} files.
In this \change{work} we approached this challenge as a red-team/blue-team exercise, and demonstrated how statistical and \change{machine-learning-based} approaches can identify faults.
While \change{the blue-team} were able to identify most outliers (i.e. defective \texttt{g-code} files), the \change{machine-learning-based} approaches had reduced accuracy, indicating that further training data is required.

Future work in this area should focus on \change{improving} and refining the selected algorithms in conjunction with an expansion of the defect methodologies. In addition, it would be interesting to determine if there are any CAD features that might effect the overall success of the defect detection strategies. %

\section*{Acknowledgment}
This work was supported in part by National Science Foundation (NSF) SaTC DGE-1931724, NSF IRES OISE-1952479, and SecureAmerica Institute.

\bibliographystyle{IEEEtran}
\bibliography{detective-gcode}
\end{document}